\documentclass[runningheads]{llncs}

\usepackage[year=2026,ID=*****]{eccv}

\newcommand{\metricblock}[1]{\includegraphics[width=.48\linewidth,
    trim=0 36 0 29, clip]{figures/metrics_3d_block_#1.pdf}}

\usepackage{eccvabbrv}

\usepackage{graphicx}
\usepackage{booktabs}

\usepackage[accsupp]{axessibility}  %

\usepackage{hyperref}

\usepackage{orcidlink}

\usepackage{eso-pic}
\usepackage{xcolor}

\makeatletter
\def\blfootnote{\gdef\@thefnmark{}\@footnotetext}
\makeatother

\usepackage[acronym]{glossaries}
\glsdisablehyper
\newacronym{mds}{MDS}{Multidimensional Scaling}
\newacronym{qi}{QI}{quasi-isodynamic}
\newacronym{mhd}{MHD}{magnetohydrodynamic}

\usepackage{microtype}
\usetikzlibrary{positioning, arrows.meta}
\usepackage[super]{nth}
\begin{document}

\AddToShipoutPictureBG*{
  \AtTextLowerLeft{
    \put(0,-25){ %
      \parbox{\textwidth}{
        \rule{0.25\textwidth}{0.4pt}\\[0.5ex] %
        \footnotesize \textcolor{darkgray}{Accepted as an extended abstract at the "Geometric Intelligence: From Vision to Scientific Discovery" Workshop, ECCV 2026.}
      }
    }
  }
}

\title{Towards Stellarator Geometry Optimisation for Nuclear Fusion}

\author{Tobias Wei{\ss}berg%
\orcidlink{0009-0005-4589-6909} \and
Moritz Heep%
\orcidlink{0009-0009-3760-8371} \and
Zorah Lähner%
\orcidlink{0000-0003-0599-094X} \and
Florian Bernard%
\orcidlink{0009-0008-1137-0003}}

\authorrunning{T.~Wei{\ss}berg et al.}

\institute{University of Bonn \&
Lamarr Institute, Germany \\
\email{weissberg@uni-bonn.de}}

\maketitle

\begin{abstract}
  We present a local geometry refinement method and a 2D latent representation that took us to the top of the ConStellaration leaderboard on the geometric task in May 2026.
\end{abstract}

\section{Background on Stellarator Optimisation}
\label{sec:intro}

Fusion power has been a central research topic since the early 1950s~\cite{post1957controlled}. However, nuclear fusion of particles requires containing plasmas that are roughly ten times hotter than the Sun's core~\cite{gerard2024stellarators}. 
One way to achieve this is to suspend the plasma in a vacuum using a toroidal magnetic field.
One such enclosure is the stellarator. Unlike its axisymmetric counterpart, the tokamak~\cite{wesson1987tokamaks}, a stellarator drives no current through the plasma to assist confinement~\cite{spitzer1958stellarator}. 
Instead, the confinement is provided by the shape of the magnetic field alone, allowing for intrinsically stable operation and rendering the design of the magnetic field a highly relevant shape-optimisation problem. %
This problem can be reduced to optimising the plasma boundary: A closed toroidal surface which, once fixed, determines the magnetic field within it~\cite{nuehrenberg1986stable}. Assessing a proposed boundary shape requires numerically solving for the \gls{mhd} equilibrium, because general three-dimensional MHD equilibria admit no known closed-form solution~\cite{hirshman1983steepest}.
\subsubsection{ConStellaration Benchmark.}
\begin{figure}[t]
    \centering
    \begin{tikzpicture}[font=\small]
        \node[inner sep=0] (elong)
            {\metricblock{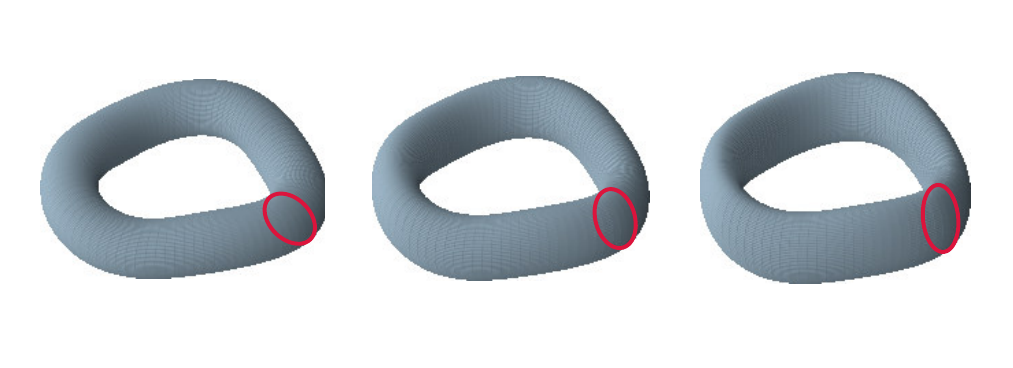}};
        \node[inner sep=0, right=3mm of elong] (ar)
            {\metricblock{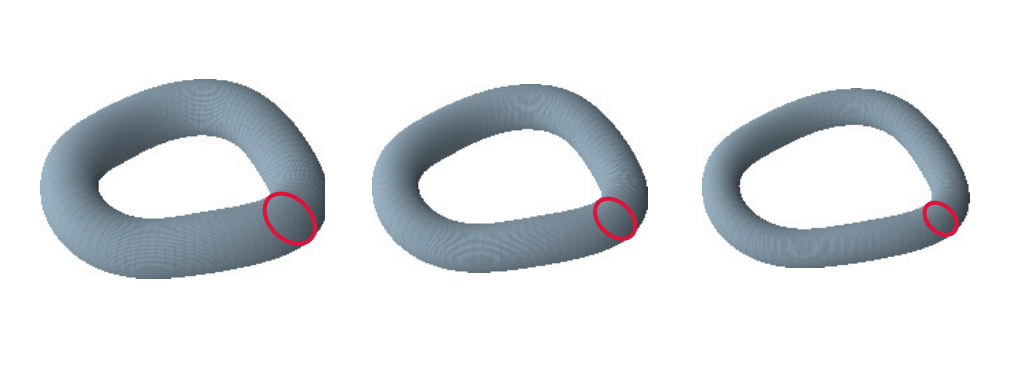}};
        \node[inner sep=0, below=5mm of elong] (tri)
            {\metricblock{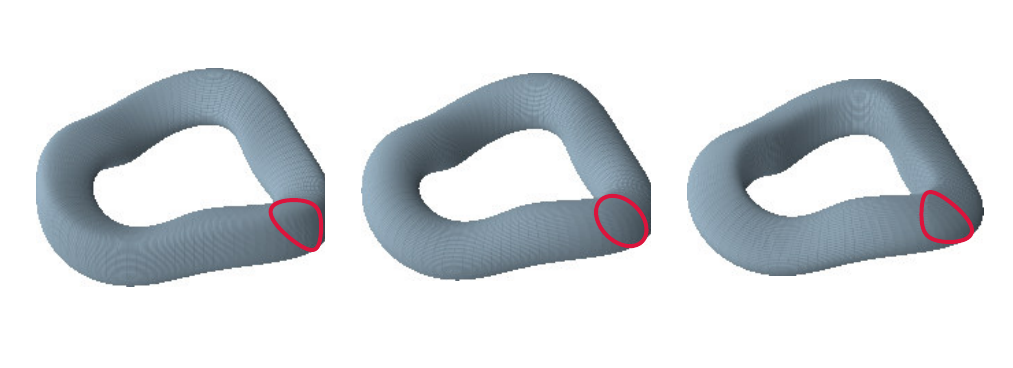}};
        \node[inner sep=0, right=3mm of tri] (rot)
            {\metricblock{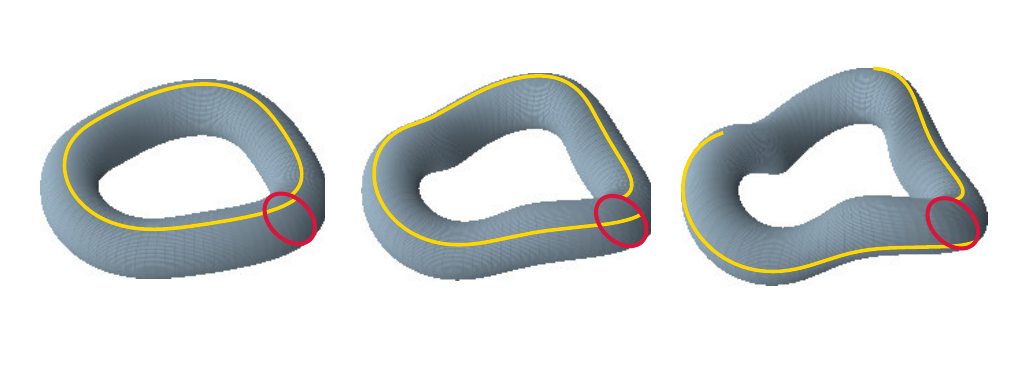}};
        \node[above=-1mm of elong] {max elongation $(\text{Obj.} \leftarrow)$};
        \node[above=-1mm of ar]    {aspect ratio $(\text{Const.} \leftarrow)$};
        \node[above=0mm of tri]   {avg. triangularity $(\text{Const.} \leftarrow)$};
        \node[above=0mm of rot]   {rotational transform $(\text{Const.} \rightarrow)$};
        \draw[-{Latex[length=2mm]}]
            ([xshift=8mm, yshift=-3mm]tri.south west) --
            ([xshift=-8mm, yshift=-3mm]tri.south east)
            node[pos=0, left]{low} node[pos=1, right]{high};
        \draw[-{Latex[length=2mm]}]
            ([xshift=8mm, yshift=-3mm]rot.south west) --
            ([xshift=-8mm, yshift=-3mm]rot.south east)
            node[pos=0, left]{low} node[pos=1, right]{high};
    \end{tikzpicture}
    \caption{Visualisation of the four metrics of the geometric task of ConStellaration. Red shows a cross-section and yellow shows a field line of the magnetic field. Arrows point toward the favourable direction: a lower objective, or the satisfying side of a constraint.}
    \label{fig:metrics}
\end{figure}
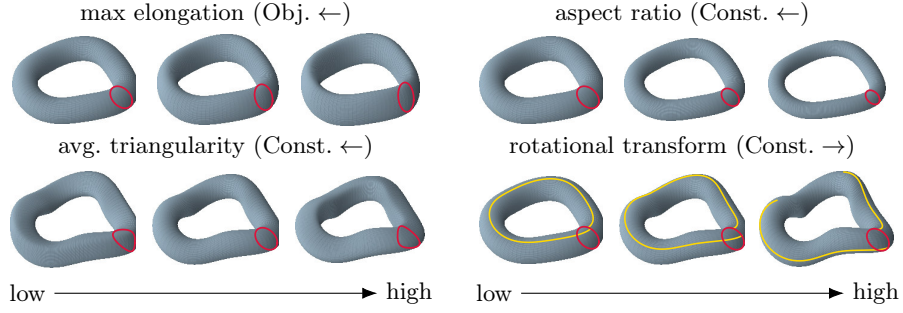

Cadena et al.\ introduced ConStellaration~\cite{cadena2025constellaration}, a dataset of ${\sim}$158k high-fidelity simulation results (using the VMEC++ simulator~\cite{hirshman1983steepest, schilling2025numerics}) for QI-like stellarator plasma boundaries, with \Gls{qi} being a sought-after physical property that promises better confinement. %
Crucially, they also established three standardised benchmark tasks, %
providing unified evaluation targets in a field traditionally characterised by bespoke and fragmented metrics.
The geometric task is the simplest %
of the three: minimise how stretched the plasma's cross-section becomes (its \emph{elongation}), while keeping the torus sufficiently compact (\emph{aspect ratio}), appropriately D-shaped (\emph{average triangularity} $\leq -0.5$), and the magnetic field lines twisted (\emph{rotational transform}). Of these four metrics, only the rotational transform requires the equilibrium; the first three are properties of the boundary surface itself. The meanings of the four metrics are further illustrated in Fig.~\ref{fig:metrics}.
We focus on this geometric task and regard the remaining tasks as a natural continuation of this work.
\subsubsection{Shape Representation.}
The domain's standard representation for toroidal surfaces are truncated double Fourier series \cite{hirshman1983steepest}: the cylindrical coordinates of the boundary are expanded in the two angles $\theta,\phi$ of the torus, $R(\theta, \phi) = \sum R_{m,n} \cos(m \theta - nN_{fp}\phi)$ and similarly $Z(\theta, \phi) = \sum Z_{m,n} \sin(m \theta - nN_{fp}\phi)$. %
This choice is long-established and well-motivated: coefficients are compact and the surface is analytically smooth. %
Furthermore, available physics simulation frameworks~\cite{schilling2025numerics, hirshman1983steepest, dudt2020desc} consume them directly, and truncation gives a natural coarse-to-fine hierarchy.
Yet, choosing the double Fourier series representation is consequential %
and largely unexamined in the light of modern geometry representations. %
For example, the Fourier representation is inherently global, as single-coefficient perturbations move the boundary everywhere, and localised geometric edits require coordinated changes across the entire spectrum.
Further, the Fourier representation hides an additional subtlety: %
it is not a representation of surfaces but of \textit{parameterised} surfaces and distinct coefficients can represent equivalent geometries. %
Spectral condensation~\cite{hirshman1985spectral} aims to alleviate this by offering a canonicalisation into a unique representation of minimal spectral width. The canonical representation, however, is defined through a non-convex minimisation and does not necessarily vary continuously with the geometry. 
\enlargethispage{\baselineskip} 
\section{Our Geometry Optimisation Method}
\begin{figure}[t]
    \centering
    \begin{subfigure}[c]{0.7\linewidth}
        \centering
        \includegraphics[width=\linewidth]{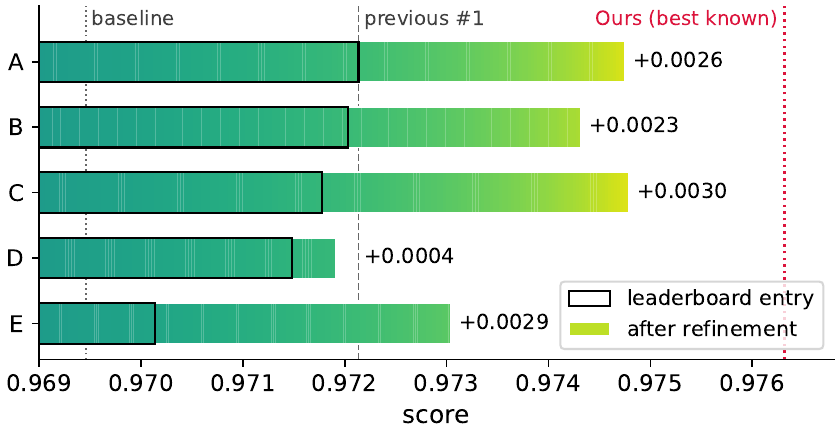}
    \end{subfigure}
    \begin{subfigure}[c]{0.22\linewidth}
        \centering
        \includegraphics[width=\linewidth]{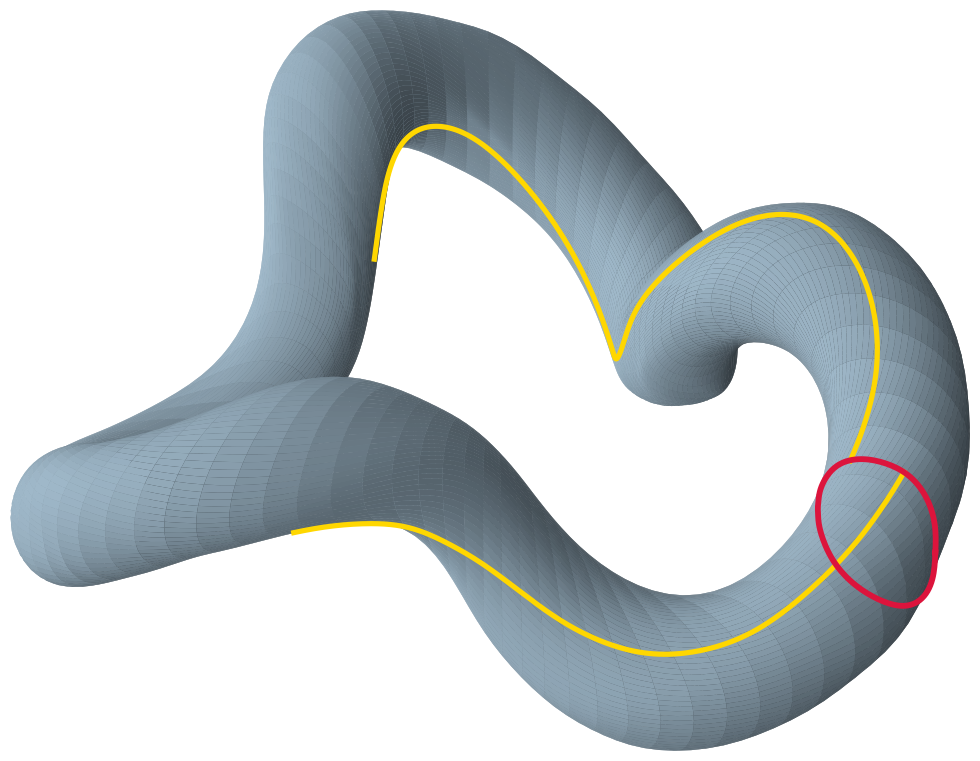}
        \vspace{2mm} \\
        {Ours \\ \hspace{2mm}(best known)}
    \end{subfigure}
    \caption{Left: Our local refinement improves each of the previous top-5 solutions. The best known solution is our own and the result of our low-dimensional parametrisation presented in the next section. %
    Right: The surface of our best known solution.
    }
    \label{fig:audit_barh}
\end{figure}

We tackle stellarator geometry optimisation as a refinement problem of existing leaderboard solutions.
To this end, we propose a simple refinement method: an augmented-Lagrangian loop following Cadena et al.~\cite{cadena2025constellaration}, but replacing the evolutionary optimisation algorithm CMA-ES~\cite{hansen2016cmaes} with local gradient based L-BFGS~\cite{liu1989lbfgs}, where gradients across the 306-dimensional input space are computed in parallel using a forward, two-point finite-difference stencil straight through the solver. With 96 concurrent VMEC++ instances per gradient this takes about one hour per optimisation run. %
Following ConStellaration~\cite{cadena2025constellaration}, we affinely transform the minimisation objective into a score value to be maximised.
The results from running our refinement for each of the previous top-5 public entries\footnote{\label{datasetdate}Leaderboard accessed on May 20, 2026; all comparisons in this work refer to this snapshot of the public leaderboard, which has since advanced.} (merging two near-duplicate submissions into one) is shown in Fig.~\ref{fig:audit_barh}. Our method improves \textit{all} of the geometries and four out of five by at least 115\% of the distance between the previous best and the previous \nth{5} solution. 
\subsubsection{2D Latent Space Model.}
We use \gls{mds}~\cite{torgerson1952multidimensional} to embed the solutions in the plane such that their pairwise Euclidean distances in coefficient space are preserved as much as possible. Given the 2D embedding positions, every point of the plane is mapped back to Fourier coefficients representing a geometry %
using mean value coordinates~\cite{floater2003mean}, evaluated with respect to the ten embedded solutions connected as a closed polygon. The resulting weights sum to one and serve as interpolation coefficients between the corresponding Fourier coefficients. This causes each solution to be reproduced exactly at its own position and makes the map smooth %
away from the vertices and well-defined on the whole plane. Additionally, it traces a curved two-dimensional slice through coefficient space rather than a plane. %

\begin{figure}[ht]
    \centering
    \begin{subfigure}{0.48\linewidth}
        \centering
        \includegraphics[trim={0 5mm 0 0}, clip, width=\linewidth]{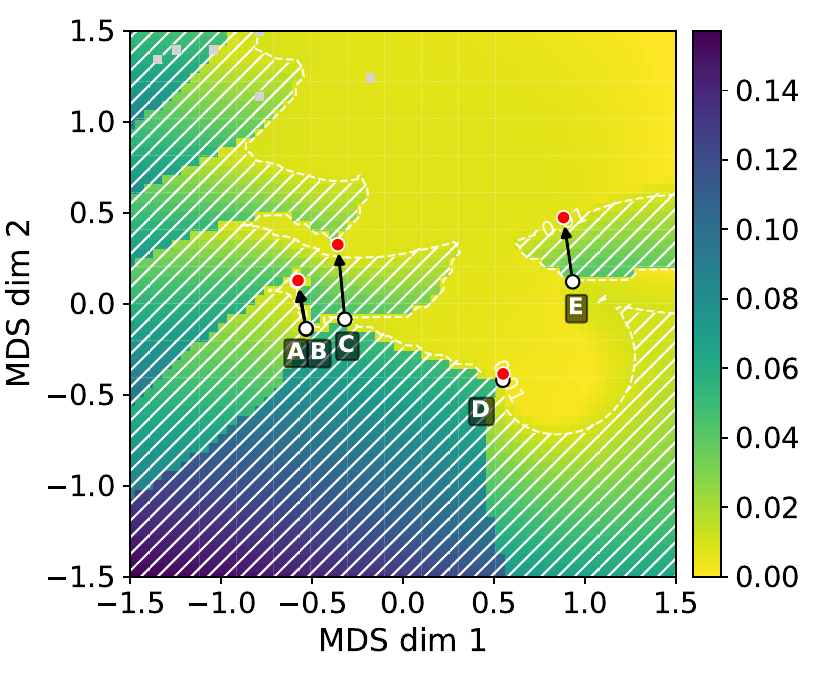}
        \caption{Max constraint violation}
        \label{fig:landscape_feas}
    \end{subfigure}
    \hfill
    \begin{subfigure}{0.48\linewidth}
        \centering
        \includegraphics[trim={0 5mm 0 0}, clip, width=\linewidth]{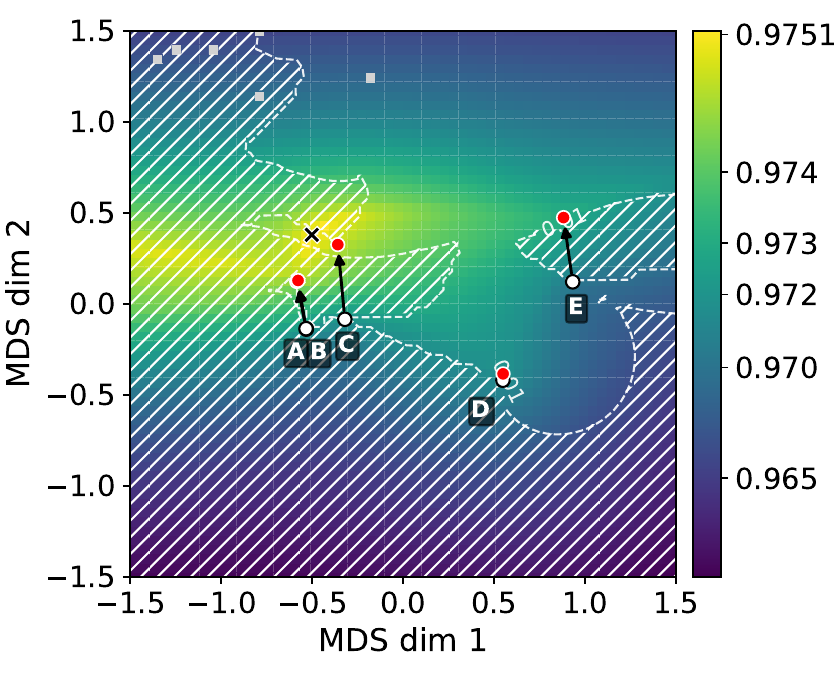}
        \caption{Objective, rescaled to score units}
        \label{fig:landscape_obj}
    \end{subfigure}
    \caption{Feasibility and objective of the latent space, where the hatched area shows constraint violations and gray color corresponds to failed evaluations. White points are the original benchmark entries, while red points show the refined solutions. %
    All 11 points shown are feasible. }
    \label{fig:landscape}
\end{figure}

We plot the loss landscape in the latent space around the previous top-5 entries and their refinements in Fig. \ref{fig:landscape}. The maximum constraint violation (Fig. \ref{fig:landscape_feas}) reveals a feasible region that is connected but not solid, with infeasible pocket intrusions, consistent with the discontinuous triangularity metric. 
The objective (Fig.~\ref{fig:landscape_obj}) varies smoothly across the slice, and its best values form a band extending beyond the refined solutions.
As the arrows indicate, our refinement can traverse infeasible regions to reach higher-scoring feasible geometries.

\subsubsection{Our Novel Solution.}
The cross in Fig.~\ref{fig:landscape_obj} marks a novel blended geometry that is feasible and scores 0.9752, exceeding every known solution\footref{datasetdate}, including all of our refinements. %
This point was found by sampling the latent, non-linear subspace generated by the top-5 leaderboard entries and our respective refinements.
The gain lies along a recombination of existing solutions, a move that line-searched local descent cannot propose.
Refining this blended design \textit{improves it further} to 0.9763, beating the benchmark by a margin more than 50\% larger than that of the original refined solutions. %

\section{Summary and Conclusion}

We proposed an elementary refinement method for the geometric ConStellaration task and a non-linear, low-dimensional latent representation based on MDS and mean-value coordinates for visualisation and sampling. Refinement improved all top-five entries, and sampling the induced subspace revealed a better geometry still, which we refined to a score of 0.9763, exceeding every entry in our snapshot. We submitted an entry scoring 0.9743, which topped the leaderboard. Limitations include that our refinement method is local by construction: it inherits the basin it starts in and requires solutions worth refining, which is why it did not reach the best known solution independently. We also restricted the evaluation to one of the three tasks. Integrating the low-dimensional representation into the optimisation to overcome basins, covering the remaining tasks, and looking for alternatives to the costly finite difference evaluations, are natural continuations.

\makeatletter
\ifeccv@review
\else
    \section*{Acknowledgements}
    Funded by German Federal Ministry of Research, Technology and Space (BMFTR) within the "Förderprogramm Fusion 2040 - Forschung auf dem Weg zum Fusions\-kraftwerk" (contract number 13F1012B).
\fi
\makeatother

\bibliographystyle{splncs04}
\bibliography{main}

@inproceedings{
cadena2025constellaration,
title={ConStellaration: A dataset of {QI}-like stellarator plasma boundaries and optimization benchmarks},
author={Santiago A Cadena and Andrea Merlo and Emanuel Laude and Alexander Bauer and Atul Agrawal and Maria Pascu and Marija Savtchouk and Lukas Bonauer and Enrico Guiraud and Stuart R. Hudson and Markus Kaiser},
booktitle={The Thirty-ninth Annual Conference on Neural Information Processing Systems Datasets and Benchmarks Track},
year={2025},
url={https://openreview.net/forum?id=NQSbGKlCpx}
}

@book{wesson1987tokamaks,
  author    = {Wesson, John and Campbell, David J.},
  title     = {Tokamaks},
  series    = {International Series of Monographs on Physics},
  volume    = {149},
  edition   = {4th},
  publisher = {Oxford University Press},
  year      = {2011},
  isbn      = {978-0199592234}
}

@book{gerard2024stellarators,
author = {Imbert-Gérard, Lise-Marie and Paul, Elizabeth J. and Wright, Adelle M.},
title = {An Introduction to Stellarators: From Magnetic Fields to Symmetries and Optimization},
publisher = {Society for Industrial and Applied Mathematics},
year = {2024},
doi = {10.1137/1.9781611978223},
address = {Philadelphia, PA},
edition   = {},
URL = {https://epubs.siam.org/doi/abs/10.1137/1.9781611978223},
eprint = {https://epubs.siam.org/doi/pdf/10.1137/1.9781611978223}
}

@article{hirshman1983steepest,
    author = {Hirshman, S. P. and Whitson, J. C.},
    title = {Steepest‐descent moment method for three‐dimensional magnetohydrodynamic equilibria},
    journal = {The Physics of Fluids},
    volume = {26},
    number = {12},
    pages = {3553-3568},
    year = {1983},
    month = {12},
    issn = {0031-9171},
    doi = {10.1063/1.864116},
    url = {https://doi.org/10.1063/1.864116},
    eprint = {https://pubs.aip.org/aip/pfl/article-pdf/26/12/3553/12590952/3553_1_online.pdf},
}

@inproceedings{schilling2025numerics,
  title={The Numerics of VMEC++},
  author={Jonathan Schilling},
  year={2025},
  url={https://api.semanticscholar.org/CorpusID:276235475}
}

@ARTICLE{post1957controlled,
  author={Post, Richard F.},
  journal={Proceedings of the IRE}, 
  title={Controlled Fusion Research-An Application of the Physics of High Temperature Plasmas}, 
  year={1957},
  volume={45},
  number={2},
  pages={134-160},
  doi={10.1109/JRPROC.1957.278361}}

@article{liu1989lbfgs,
  title={On the limited memory BFGS method for large scale optimization},
  author={Dong C. Liu and Jorge Nocedal},
  journal={Mathematical Programming},
  year={1989},
  volume={45},
  pages={503-528},
  url={https://api.semanticscholar.org/CorpusID:5681609}
}

@article{hirshman1985spectral,
  title={A convergent spectral representation for three‐dimensional inverse magnetohydrodynamic equilibria},
  author={Steven P. Hirshman and Harold Weitzner},
  journal={Physics of Fluids},
  year={1985},
  volume={28},
  pages={1207-1209},
  url={https://api.semanticscholar.org/CorpusID:122226875}
}

@article{floater2003mean,
title = {Mean value coordinates},
journal = {Computer Aided Geometric Design},
volume = {20},
number = {1},
pages = {19-27},
year = {2003},
issn = {0167-8396},
doi = {10.1016/S0167-8396(03)00002-5},
url = {https://www.sciencedirect.com/science/article/pii/S0167839603000025},
author = {Michael S. Floater}
}

@article{torgerson1952multidimensional, title={Multidimensional Scaling: I. Theory and Method}, volume={17}, DOI={10.1007/BF02288916}, number={4}, journal={Psychometrika}, author={Torgerson, Warren S.}, year={1952}, pages={401–419}}

@article{spitzer1958stellarator,
  title   = {The Stellarator Concept},
  author  = {Spitzer, Jr., Lyman},
  journal = {Physics of Fluids},
  year    = {1958},
  volume  = {1},
  number  = {4},
  pages   = {253--264},
  doi     = {10.1063/1.1705883}
}

@article{hansen2016cmaes,
  title={The CMA Evolution Strategy: A Tutorial},
  author={Nikolaus Hansen},
  journal={ArXiv},
  year={2016},
  volume={abs/1604.00772},
  url={https://api.semanticscholar.org/CorpusID:15038271}
}

@article{nuehrenberg1986stable,
title = {Stable stellarators with medium {$\beta$} and aspect ratio},
journal = {Physics Letters A},
volume = {114},
number = {3},
pages = {129-132},
year = {1986},
issn = {0375-9601},
doi = {10.1016/0375-9601(86)90539-6},
url = {https://www.sciencedirect.com/science/article/pii/0375960186905396},
author = {J. Nührenberg and R. Zille}
}

@ARTICLE{dudt2020desc,
       author = {{Dudt}, D.~W. and {Kolemen}, E.},
        title = "{DESC: A stellarator equilibrium solver}",
      journal = {Physics of Plasmas},
         year = 2020,
        month = oct,
       volume = {27},
       number = {10},
          eid = {102513},
        pages = {102513},
          doi = {10.1063/5.0020743},
       adsurl = {https://ui.adsabs.harvard.edu/abs/2020PhPl...27j2513D}
}
\end{document}